\documentclass[final,5p,times,twocolumn]{elsarticle}
\usepackage{amssymb}
\usepackage{float}
\usepackage{graphicx,latexsym}
\usepackage{amsmath}
\usepackage{color}

\begin{document}

\title{Distribution of plates sizes tell the thermal history in a simulated martensitic-like phase transition}
\author{F. \c Tolea, M. \c Tolea*, M. Sofronie, M. V\u aleanu}

\address{
 National Institute of Materials Physics,
POB MG-7, 77125
 Bucharest-Magurele, Romania.}

\begin{abstract}

A phenomenological 2D model, simulating the martensitic transformation, is built upon existing experimental observations that the size of the formed plates -in direct transformation- decreases as the temperature is lowered; then they transform back in reversed order. As such, if a reverse transformation is incomplete ("arrested"), the subsequent direct one will show anomalously large number of big size plates-old plus newly formed- but consequentially a depletion of intermediate sizes, due to geometrical constraints, phenomenon that generates thermal memory.

\vskip 0.5cm
\begin{it}
\end{it}

\end{abstract}

\maketitle

\section{Introduction}

The temperature memory effect, as manifested by shape memory alloys, remarkably shows that a thermodynamic system can remember not only the sense of a transformation, as in magnetic hysteresis, but also quasi-precise temperatures at which certain actions were performed in the past, such as the interruption of a previous phase transition. Naturally, the subject attracted a number of experimental efforts (see, e.g. \cite{Mad-Scripta1,Mad-Scripta2,Wanga,JALCOM1, JALCOM2,Wang_ML,Wang-IJSNM,Airoldi1,R-AJAP,R-AACTA}  towards a clear description, but surprisingly few theoretical scenarios have been proposed (see e.g. \cite{Mad-Scripta2,Airoldi1,R-AJAP} ), an unanimously accepted phenomenological model being so far rather elusive.

	Succinctly, the temperature memory effect (also called "thermal arrest") consists in the system remembering one -or several- previous incomplete phase transition(s), as described in the following four-steps process: at step I, a complete direct phase transition (parent to product) is performed, then at step II the reverse (product to parent) transformation is "arrested" -stopped before completion- at a temperature "$T_A$". Next, at step III, another complete direct transformation is performed and finally at step IV a complete reverse one. In this final reverse transformation, the calorimetric signal shows a dip at a temperature close to $T_A$, at which the transformation was previously arrested at step II.

	Efforts are on-going towards understanding of the thermal memory effect. Previous theoretical works suggest that the release of stress from the untransformed martensite can increase its transformation back temperature, which results in a separated second peak \cite{Mad-Scripta2,Wang-IJSNM,Airoldi1}. J.Rodriguez-Aseguinolaza et al.  \cite{R-AJAP} propose a more quantitative model also relaying on the redistribution of stress amongst martensite fractions and the parent matrix after one or multiple incomplete reverse transformations.

	In this paper we propose a different mechanism possibly relevant for the thermal memory effect, namely that incomplete thermal circles influence the distribution of the product phase plates sizes, making the latter a "witness" of the thermal history - which should be readable by a calorimetric scan.

	The idea is rather simple and is based on the existing microscopic evidence that the martensitic transformation takes place with formation of plates, which do not grow indefinitely, but rather stop at a certain maximum size; then, the phase transition proceeds by formation of other plates. There is also evidence that this intrinsic maximum size decreases with temperature (see, e.g. \cite{Jost}).  Keeping in mind that the size of a formed plate possesses such an intrinsic limitation, it is a simple next step to assume that geometrical constraints may impose supplementary limitations: the new plates can be limited in growth by the existing puzzle of neighboring plates. As such, the distribution of plates sizes becomes history dependent and it will be shown that this property can generate thermal memory.

	The outline of the paper is as follows:  in Section 2, the general assumptions of the model are described, then a numerical illustration is given in Sections 3 and 4; Section 5 discusses briefly the relevance and limitations of the calculations, and Section 5 concludes the paper.

\section{Model}

In this section we present the basic assumptions within a simplified phenomenological model for a solid state phase transition inspired by certain properties of the martensitic transformation. The basic features of the proposed model are:

\begin{description}
  \item[(i)] {\it Finite size plates.} The product phase, once nucleated,  grows in form of plates only up to some intrinsic maximum size where the growth stops. Then, the phase transition proceeds by formation of new plates (rather than further growth of existing ones). In the numerical simulations, this intrinsic maximum size will also be considered to decrease with temperature \cite{Property}.
  \item[(ii)]{\it Key role of geometrical constraints.} The plates may stop growing sooner if they encounter other -already existing-  plates which get in the way of their growth. The geometrical constrains thus work in one sense, of reducing the size of certain plates, and in this way influencing the plates distribution statistics.
  \item[(iii)]{\it Reverse transformation in reversed sizes order.} In the reverse transformation, the plates return to the parent phase in the reverse order of their formation, i.e. the smaller plates which were the last to form are the first to transform back and the largest plates transform back last.
\end{description}

	The property (i) already indicates a structured product phase composed of plates of different sizes, which can in principle enclose information.

 Property (ii) imposes a dependence on the specific sample history meaning that the number and sizes distribution of preexisting plates statistically influence the sizes of newly formed plates. This is the proposed key mechanism responsible for the thermal memory effect.

  The property (iii) describes the reverse transformation. Its justification relies in the higher surface-to-volume ratio for smaller plates which makes them the first to become thermodynamically unstable (a numerical exemplification will be given in Section 4).

	The above described phase transition is schematically illustrated in Fig.1. On the upper row, a direct transformation is indicated, in which the larger "martensite" plates nucleate and grow first (the red squares in panel 2), then as temperature decrease, intermediate plates form (yellow squares in panel 3), and finally the smaller blue squares form at lowest temperatures (panel 4).

	Let us now assume that in the reverse transformation martensite plates transform back in reversed order, starting the smallest blue squares and followed by the yellow ones (4-2 reverse sequences). Now if the reverse transformation is stopped ("arrested") before the bigger red plates can transform back, they will remain untransformed and a subsequent direct transformation will start from this pre-existing plates distribution, as depicted in Fig.1, second row (panels 2-5-7), and one can notice the "anomalously" large number of big size plate (those remained untransformed and the newly formed ones). A first direct effect is seen in panel 6, corresponding to temperatures favoring the yellow intermediate plates,  but which now cannot grow to their normal size because they cannot fit in the existing puzzle; instead the formed germs grow only to blue sizes. The final sizes distribution is considerably changed.

The limitation of plates growth, their random positioning in the parent matrix, and the geometrical constrains in growth are non-equilibrium effects, reflecting the pronounced metastable character of the transformation.

\begin{figure}[ht]
\centering
\vskip -1.5cm
\hskip -2cm
\includegraphics[scale=0.37]{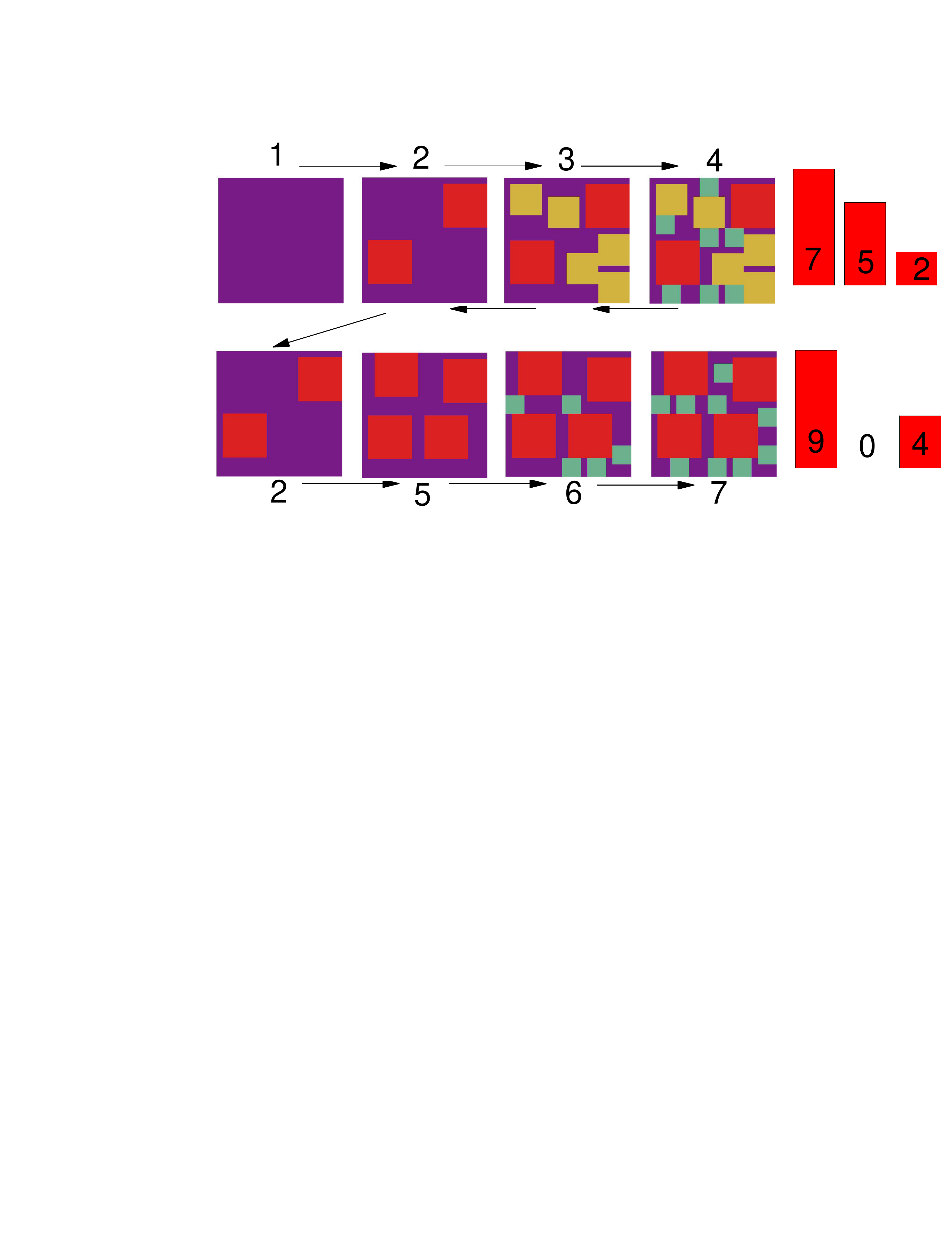}
\vskip -8cm
\caption{Schematic representation of thermal memory encrypted in the distribution of plates sizes.  The main assumptions made are that the sizes of the formed plates decrease with temperature ($T_{1}>T_{2(5)}>T_{3(6)}>T_{4(7)}$), and they transform back in reverse order of their formation. A first direct transformation ($1\rightarrow 4$) is followed by an incomplete reverse one ($4\rightarrow 2$) and a second direct transformation ($2\rightarrow 7$). Sizes distribution for the panels 4 and 7 are indicated by the bar charts. In the ($2\rightarrow 7$) transformation, intermediate size plats (yellow) are missing, because they cannot fit in the puzzle defined by the "anomalous" large number of bigger (red) plates, therefore they only grow to a "blue" size.}
\end{figure}

The intuitive scenario depicted in Fig.1 shall be numerically tested in the following section.

\section{Numerical simulations assuming isothermal direct transformation.}

In our numerical simulation, the product phase will be assumed to nucleate in random positions in the parent matrix and to grow in the form of squares which gradually fill the  available surface. If a square, during its growing process, encounters neighboring (already existing) squares, or the system borders, it will stop growing, remaining at that respective size.

While the 2D model is justified by the plates-like growth of martensite germs (also, the alloys can be obtained as thin films, etc.),  choosing a square shape for the plates should be regarded as a simplifying assumption- with the goal to just offer an illustration of  the phenomenology.

	It is a good point to mention that an interesting debate is still on-going whether the martensite transformation is isothermal or athermal \cite{PRL1997, PRL2001,pss-a,Intermet_mag,SSC1,SSC2}, the balance seemingly leaning towards the isothermal side, even if sometimes with an atypical long transformation time \cite{Scripta_Iso,Scripta_Iso_Lee, Scripta_Salas}. We shall assume the  simplest formula for a isothermal nucleation rate (denoted $J_{A\rightarrow M}$), governed by the activation critical free energy $\Delta\Omega_c(T)$ (corresponding to the formation of the critical cluster)\cite{N1,N2}:

\begin{equation}
J_{A\rightarrow M}(T)=J_0 S ~~ exp \Bigg ( -\frac{\Delta\Omega_c(T)}{k_BT}\Bigg )
\end{equation}

For our 2D case, the free energy gain when a product phase germ is formed (which in our model is a square of side L) can be written: $\Delta\Omega_{A\rightarrow M}=L^2\cdot \Delta F+4L\cdot\sigma$.The condition $\frac{\partial}{\partial L}\Delta\Omega_{A\rightarrow M}=0$ yields the critical square side $L_c=-2\sigma / \Delta F$, from which it results: $\Delta\Omega_c(T)=-4\sigma^2/\Delta F$, $\Delta F=\epsilon (T-T_0)$ being the free energy difference in the linear approximation. $T_0$ is the equilibrium transition temperature; $\epsilon$ and $\sigma$ are respectively the entropy difference (between parent and martensite phase) and the interface energy per unit length.
The interface energy is always positive and may be a sum of elastic and inelastic contributions. The factor S in Eq.1 is the untransformed remained surface (the initial total surface is considered $A^2$ ). $J_0$ is a proportionality constant, its value choice for the numerical simulations being described below.

The quantities introduced so far are typical for a first order phase transition, but now we need to impose  the supplementary specific property that the formed germs have an intrinsic limitation in growth- moreover, their final size also decreases as the temperature is lowered.  One possible way to introduce intrinsic limitation in growth is to assume a proportionality between the size of the critical germ ($L_c$) and the size of the final plate $(L_f)$ :$L_f=nL_c$ [meaning $L_f\sim 1/(T-T_0)$]. One can notice that the size $L_c$ of the critical germ decreases with decreasing temperature below $T_0$, then it results from our proportionality hypothesis, that the final size of the plates decrease as well with temperature. The phenomenologically introduced limitation in the plates growth completes our model for the direct transformation for which numerical results are presented in this section.

In order to proceed with the real-time numerical simulation of the phase transition, one needs to describe also the heat exchange with the calorimeter, which is considered simply: $dQ/dt=\beta A^2 (T-T')$,
where $T$ and $T'$ are, respectively the sample and thermostat temperatures and $A^2$ is the total sample surface. In the numerical simulations, the thermostat temperature will vary  linearly, the typical case in DSC experiments: (say, $T'=a∙t$, with $a<0$ for the direct transformation). If the phase transition is initiated, part of the transferred heat is used as latent heat ($Q_L$), the rest warms (or cools) the sample with the amount per unit time: $dT/dt=d(Q-Q_L)/dt \cdot 1/(cA^2)$  , where c is the specific heat. $Q_L$ is calculated as follows: the germs formation ratio is given by Eq.1, then each germ has a size dependent growing time, let this time be denoted $t(L)$; then $dQ_L/dt=\sum 1/[t(L)] \cdot L^2 D_E$, the summation being over all growing germs, $D_E$ being the latent heat per surface unit. In the numerical calculations the border energy will also be considered.

A brief discussion is in order about the parameters used in the numerical simulation. Since our intention is a purely qualitative insight on the phenomenon, the parameters are adimensional, as used in the numerical simulation. First, the sample is modeled by a square of $120 X 120$ (discrete points), Much like the finite differences method for solving differential equations, the discretization brings an important advantage by reducing the computing time, with the draw-back that the continuous space is just approximately described.  The product phase plates are squares with the side ranging from $L_{Max}=12$ points, to $L_{Min}=3$.

The equilibrium transition temperature is generically chosen as $T_0=300$. The parameters appearing in the exponential from Eq.1 are chosen such as $4\sigma^2/(\epsilon k_B T_0 )\sim 700$ , $J_0=0.1$, which lead to a "martensite start" temperature of approx $T_{MS}=T_0-15$. Technically, $T_{MS}$ can be calculated from the condition that the temporal integration of the nucleation rate reaches the value $1$, corresponding to the formation of the first germ $\int_{t=0}^{t_{MS}} J_{A\rightarrow M} (T(t) )dt=1$. Other parameters (defined in this section) were chosen such that $(D_E L_{Max}^2)/t(L_{Max})< \beta A^2 (T_0-T_{MS})$, meaning that the latent heat is in general less that the total heat exchanged with the reservoir so that the direct phase transition, for instance, is accompanied by a permanent cooling of the sample, making possible the formation of progressively smaller plates.

\begin{center}
\begin{figure*}[ht]
\includegraphics[scale=0.45]{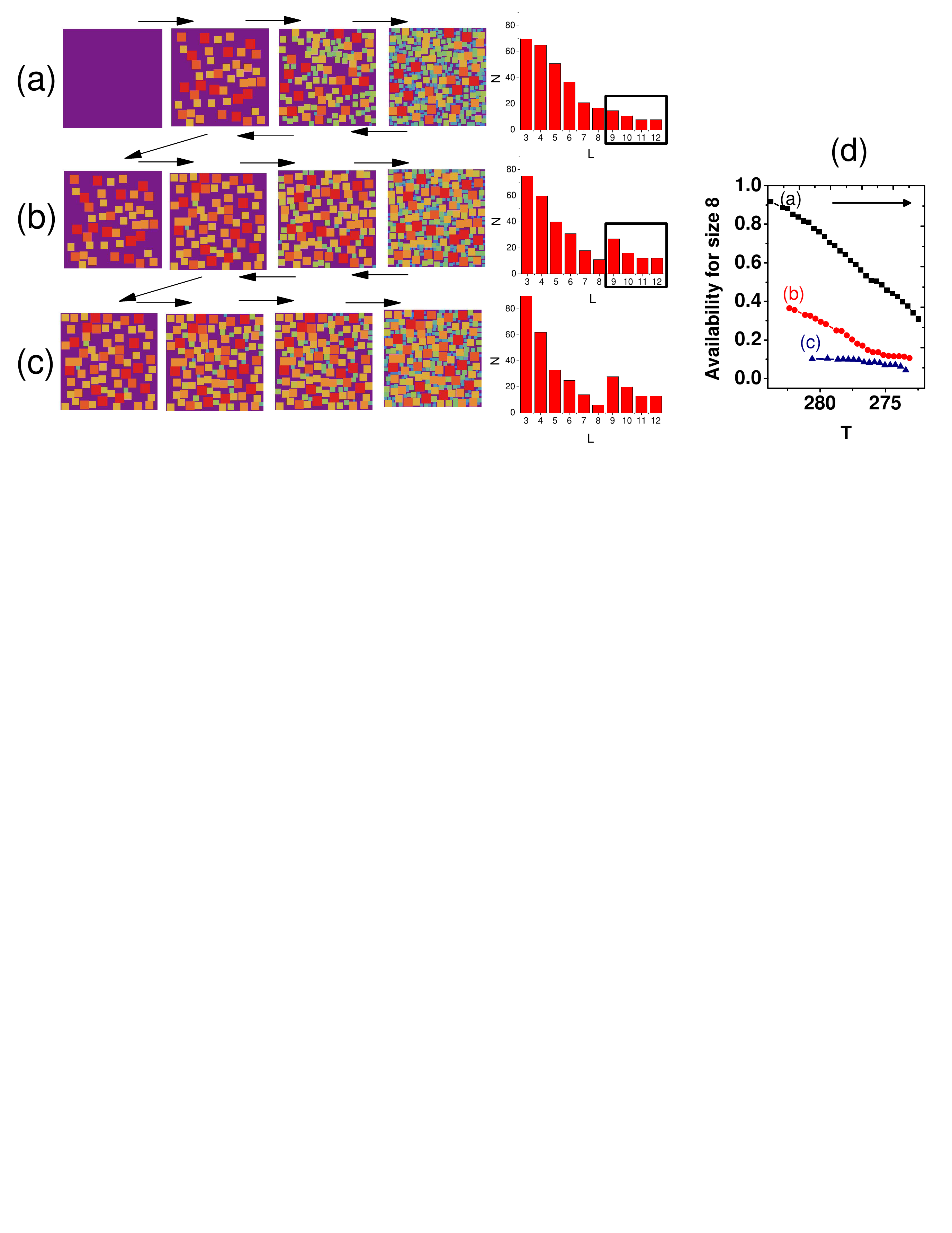}
\vskip -10.5cm
\hskip 3cm
\caption{(a) Left to right: a direct transformation resulting in the plates distribution sizes plotted in the bar chart. Next an incomplete reverse transformation is performed leaving the plates with sizes bigger than $8$ untransformed -indicated with a rectangle in the bar chart-. (b) A subsequent direct transformation leads to a different
plates distribution with increased number of big plates (sizes 9 to 12) and a depletion of intermediate sizes (sizes 7 and 8) . (c) Repeating the arrest procedure a second time accentuates the sizes unbalance. (d) Relative number of places where a square of size 8 can fit, for the three situations (a)-(c), as the temperature is lowered in the direct transformation (see description in text). Upper arrow indicates the transformation sense.}
\end{figure*}
\end{center}

The first plates to form are also the largest, and in our model the maximum size will be $L_{Max}=12$, from which the proportionality constant assumed between the critical germ and the final plate size can be derived. Technically, an initially formed germ will be considered of size $2 X 2$, randomly placed on the untransformed surface. Then it will "grow" to the final plate size, unless it encounters neighboring plates or the system borders, case in which it only grows to the maximum size geometrically possible.

Let us now analyze the consecutive transformations  schematically depicted in Fig.2 (a)-(c), which is just a more sophisticated version of Fig.1. On the first row, from left to right a direct transformation is simulated which consists in formation of plates with decreasing sizes as temperature is lowered, finally resulting in the sizes distribution plotted in the bar chart. The distribution is balanced, in the sense that the areas of the product phase being in the form of different sizes squares is comparable (for instance there are 6 plates of side 12 with a total area of 864, and 70 plates of side 3 with a total area of 630, etc.).

The sizes distribution presented in the bar chart Fig.2a can for instance be compared with the chart distributions given in \cite{Ghosh}, to which a resemblance can be noticed for big size plates. However, for small size plates \cite{Ghosh} gives a decrease not captured in our simulation most likely because: (i) \cite{Ghosh} addresses incipient stages of the phase transition (usually few percents transformed, and mostly 30\%) and (ii) the smallest plates size in our model was imposed by the discretization used,  so the number of plates with the size "3" should enclose also smaller plates from the physical situation. The size distribution of austenite nano-crystals is discussed in, e.g. \cite{Glezer}.

After the direct phase transition was simulated, we address the reverse transformation which is initiated with the smaller plates being the first to transform back and the system going basically through the same configurations in reverse order. In order to simulate the thermal arrest, this reverse transformation is incomplete. The sizes of the plates remained untransformed are shown in a rectangle. They define the configuration from which the next direct transformation will start (Fig.2b). One can notice an anomalously large number of big plates, consisting of the mentioned plates (remained untransformed) plus newly formed ones. An important consequence of this will be a depletion in the number of intermediate sizes plate, which can now fit with difficulty in the puzzle set by the large number of big plates. As such, most plates will grow to a lower size than the one maximum allowed at a given temperature. Repeating the "arrest"  procedure enhances the effect (known as "hammer effect"\cite{Airoldi1,R-AJAP}), and this situation is shown in Fig.2c.

The physical mechanism for depletion of intermediate sizes is the one schematically depicted in Fig.1, originating in the increased role of geometrical constraints in the direct transformations following a thermal arrest. We present in Fig.2d the temperature dependence of the relative number of available places to fit a size $8$ square for the transformations (a)-(c). This number is calculated by dividing the number of available spots to fit a size $8$ plate with the number of places for the size $3$ plates (smallest in our model). At the ignition of the initial transformation (upper black curve) the ratio is nearly $1$, as all sizes can equally fit everywhere, except for near the system borders. As the transformation proceeds, smallest plates (size $3$) fit significantly easier than larger plates (size $8$) and the ratio decreases. After the arrest procedure, size $8$ plates fit even more difficulty, which will determine them to grow to a smaller size, similar to the scheme shown in Fig.1. For all temperatures, the number of available places for the size $8$ plates in transformations (b) and (c) are much reduced as compared to transformation (a), leading to the depletion seen in the bar charts. It is interesting to notice that the (a), (b) and (c) lines in Fig.2d start at progressively lower MS temperature, suggesting that that the thermal arrest may slightly increase the temperature hysteresis.

In this section, a number of assumptions have been made regarding the direct transformation, however regarding the reverse transformation it was only assumed that the plates transform back in the reverse order of their sizes.  A possible transformation law example will be given in the next section.

\section{Reading thermal memory by reverse transformation (toy model illustration)}

Performing reverse transformations is a key part in the arrest experiments. In this section we assume a phenomenological transformation rate to simulate DSC curves for the reverse transformation.

\begin{center}
\begin{figure*}[t]
\vskip -2cm
\includegraphics[scale=0.54]{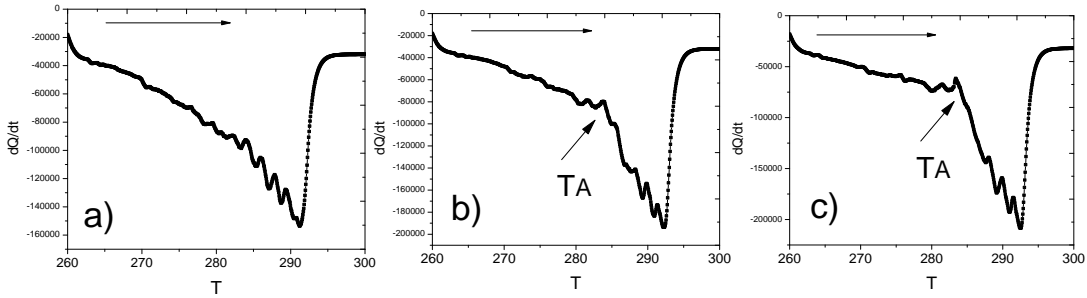}
\vskip -7.5cm
\caption{ Simulated calorimetric signal (in arbitrary units) of the reverse transformation after previous complete cycles (a), after one previous thermal arrest at the temperature "$T_A$" (b) and after two successive arrests at the same temperature (c).}
\end{figure*}
\end{center}

The simplest way to introduce the property (iii) is to assume that the interface energy has an elastic component, which can actually be responsible for initiation of the reverse product-parent transition. Let us denote the elastic fraction of the border energy with  $4L∙\sigma '$ (necessarily, one must have $\sigma '\leq \sigma$).  This negative elastic energy that can be restored to the system will gradually make the martensite plates become energetically unfavorable, at a certain temperature, that is size dependent, and can be easily calculated to be:

\begin{equation}
T_L=T_0-\frac{4\sigma '}{\epsilon L}
\end{equation}

Note that, indeed, the above formula implies automatically that the smaller martensite plates become unstable first (at lower temperatures) and will therefore be the first to transform back into austenite. Moreover, the reverse transformation is initiated below $T_0$, and possibly completed also below $T_0$. (For a discussion of this possibility that the $M\rightarrow A$ transformation is initiated at  $A_S<T_0$, due to the stored elastic stress, see for instance \cite{OP1}).
Once the condition Eq.2 is achieved for a given size plates, they begin to transform back with a certain transformation rate, which can be generically expressed:

\begin{equation}
J_{M\rightarrow A}(L,T)=J_0 ' N_L ~~\theta (T-T_L)f(L,T),
\end{equation}

where $J_0 '$ is a constant, $N_L$ is the number of existing martensite squares of side length L and $\theta (x)=1$ for $x>0$, and $0$ othewise. The most simple choice for $f(L,T)$ may be to consider it constant, alternatively one may use exponential form such as $f(L,T)=exp [ -\frac{\Delta\Omega(L,T)}{k_BT}]$,
with $\Delta\Omega(L,T)=-\epsilon L^2 (T-T_0 )-4\sigma ' L$, as we employed for the sake of a graphical illustration in Fig.3. We stress again that Eq.3 fulfils the property (iii)- from Section 2, however we do not claim any microscopic justification for it, nor for the particular form of $f$.
 An important difference between Eq.3 and Eq.1 is that the $M\rightarrow A$ reverse transformation can actually stop if there exists no thermodynamically unstable squares of a given side L, and the transformation of the bigger squares cannot start until the temperature is further raised and the condition in  Eq.2 is fulfilled.

The numerical results are reflected in Fig.3, panel a) corresponding to a reverse transformation following  complete previous transformations. Panel b) corresponds to a reverse transformation following an incomplete one (which was arrested at the temperature "$T_A$") and a complete direct transformation. Panel c) illustrates the "hammer" effect, with a more pronounced dip after the arrest sequence is repeated.  The plotted curves show also a number of oscillations, not related to the thermal memory - the small local dips corresponding to the plates of a certain size being transformed almost completely while the plates immediately smaller did not start the transformation.  This is a limitation of the model, as we considered only $10$ different sizes for the plates, a better discretization being perhaps necessary for quantitatively accurate plots.

\section{Relevance and limitations of the model}

 A number of experimental papers (see, e.g. \cite{R-AACTA,Jost}) present microscopic evidence that, when increasing the temperature, the martensite plates transform back to austenite in reversed order of their formation.  More general, even if some alloys undergo morphologic changes  during cooling, the reverse transformation takes places through the same stages as the direct one - but in reverse order \cite{Cai}. Also in the paper \cite{Jost}, chosen as example, the author comments on the fact that during the direct transformation, larger martensite plates are formed first -at higher temperatures- and smaller plates subsequently -at lower temperatures- . These aspects justify our assumptions described in Section 2. However, they are not universal features valid for all martensitic transformations. Also, one can in principle argue that the fact that smaller plates form at lower temperatures, when the phase transformation is in an advanced stage, may in principle be exclusively due to geometrical constrictions and/or accumulated stress in the sample, rather than an intrinsic thermal property, as assumed in our numerical simulations.

	Further limitations of our numerical approach reside in the simplified shape choice (squares) of the plates, so more involved numerical approaches (in future works) may assume more realistic lamellar geometries and finite thickness.

	 Even more drastically are the simplifications regarding the nucleation rates. While, for instance, it seems justified to assume isothermal martensitic transformation \cite{PRL2001}, the solid-solid nucleation rates are more complex then the simplest Eq.1 assumed, with possible important consequences. Very recent state-of-the-art simulations actually suggest that solid-solid phase transitions take place in a two-step process, with an intermediate liquid-like phase \cite{Nature}.

	Nevertheless, our intention here is just to give plausibility arguments for a mechanism potentially relevant for the temperature memory effect.

\section{Conclusions}

We propose a simple phenomenological model for a solid state phase transition, inspired by some experimentally observed features of the martensitic transformation. Then it is shown than the model naturally exhibits thermal memory.

	First, the key assumptions are described, namely that the direct phase transition takes place by formation of finite plates, whose seizes are temperature dependent (with bigger plates formed first) and also -possibly- geometrically restricted by existing surrounding plates puzzle. Secondly, numerical simulations are performed considering isothermal nucleation rates for the direct transformation and the assumption that the reverse transformation takes place in reverse order (i.e. smaller plates transform back first).

	Based on the described model, if a reverse transformation is incomplete (arrested), the larger plates will remain untransformed and a subsequent direct transformation will show an anomalously large number of big sizes plates, consisting of the untransformed ones plus the newly formed. As temperature is further decreased, intermediate size plates may encounter difficulties to geometrically fit - which results in them growing to a smaller final size and a general depletion of the sample of intermediate sizes. The effect is accentuated if the arrest scheme is repeated (the hammer effect). The dependence of the plates sizes distribution on the thermal history is a memory effect. Finally, a numerical model is proposed also for the reverse transformation, showing that the memory effect should be "readable" by a calorimetric signal.

	In a sense our approach has similarities with \cite{R-AJAP,R-AACTA}, the size of the plates playing the role of stress density, and each new direct transformation has a "stress free" start - or in our model big plates are formed first. However, the supplementary key role of geometrical constrictions imposed by the surrounding plates is a distinct feature.

	We hope that our phenomenological model may motivate further theoretical efforts to better understand the thermal memory effect, and also new experiments in particular seeking a correlation between incomplete thermal circles and the distribution of plates sizes.

\section{Acknowledgements}

We acknowledge support from the Romanian Ministry of National Education,  grants PN-II-ID-PCE-2012-4-0516 and
Core program 45N/2009.

\vskip 1cm
*corresponding author e-mail: tzolea@infim.ro


\begin{thebibliography}{99}
\bibitem{Property} The size of the critical nucleation germ is decreasing with temperature so this can suggest a decrease also of final plate size. Growth of initial germ is also hampered at lower temperatures by increase "stiffness" of surrounding atoms. There is also experimental evidence to sustain this property.
\end{thebibliography}
\end{document}